\documentclass[
aps,%
10pt,%
final,
notitlepage,%
oneside,%
twocolumn,%
nobibnotes,%
nofootinbib,%
preprintnumbers,%
superscriptaddress,%
showpacs,%
centertags,%
showkeys,%
amsmath,%
amssymb]{revtex4}

\usepackage{amsmath}
\usepackage{amssymb}
\usepackage{graphicx}

\newcommand{\vecc}[1]{\mbox{\boldmath $#1$}}
\newcommand\m[1]{\mathrm {#1}}
\newcommand\nn{\nonumber}

\def\dd{\mathrm d}

\def\phi{\varphi}
\def\vare{\varepsilon}

\begin{document}

\title{Recoil proton distribution in high energy photoproduction
processes}

\author{E.~Barto\v{s}}
\affiliation{Institute of Physics, Slovak Academy of Sciences,
Bratislava}

\author{E.~A.~Kuraev}
\affiliation{Joint Institute for Nuclear Research, Dubna, Russia}

\author{Yu.~P.~Peresunko}
\affiliation{NSC KIPT, Kharkov, Ukraine}

\author{E.~A.~Vinokurov}
\affiliation{NSC KIPT, Kharkov, Ukraine}

\date{\today}

\begin{abstract}
For high energy linearly polarized photon--proton scattering we have
calculated the azimuthal and polar angle distributions in inclusive
on recoil proton experimental setup. We have taken into account the
production of lepton and pseudoscalar meson charged pairs. The
typical values of cross sections are of order of hundreds of
picobarn. The size of polarization effects are of order of several
percents. The results are generalized for the case of
electroproduction processes on the proton at rest and for high
energy proton production process on resting proton.
\end{abstract}


\maketitle

We have considered below the experimental setup of processes of
charged pair $a_-a_+$ production (pseudoscalars, leptons) by high
energy photon scattering on proton at rest frame with following
detection of recoil proton
\begin{gather}  \label{eq1}
\gamma(k,\vare)+\m{p}(p)\to a_-(q_-)+a_+(q_+)+\m{p}(p'),\nn\\
s=2k.p,\;k^2=0,\;p^2=(p')^2=M^2,\;q_-^2=q_+^2=m^2.
\end{gather}

Two different mechanisms of pair production must be considered. One
corresponds to the pair creation by two photons Bethe--Heitler (BH).
Another one is the bremsstrahlung (B), which corresponds to the case
when pair is created by single virtual photon (we have implied the
lowest in QED coupling constant $\alpha=1/137$ contributions). The
contribution of B type is suppressed compared with one of BH type by
factor $|q^2|/s$. As for interference of B and BH amplitudes it is
exactly zero in the inclusive on recoil proton setup we have
considered below.

The accuracy of formulae given below are determined by the terms we
have omitted systematically compared with terms of order of unity
\begin{gather}  \label{eq2}
1+O(\frac{\alpha}{\pi},\frac{|Q^2|}{s},
\frac{m^2}{s},\frac{M^2}{s}),\quad Q=p-p'.
\end{gather}

In the peripheric kinematical region $ s\gg|q^2|\sim M^2,$
effectively works the Infinite momentum Frame (IMF) or Sudakov
\cite{BFKK} parametrization of transferred momentum and the
4--momenta of final particles
\begin{gather}  \label{eq3}
Q=\alpha_q\tilde{p}+\beta_q k+q_\bot,\quad q_\pm=\alpha_\pm
\tilde{p} +x_\pm k+q_{\pm\bot},\\ c_\bot p=c_\bot k=0,\quad
\tilde{p}=p-k\frac{M^2}{s},\nn\\ \tilde{p}^2=0,\quad
q_\bot^2=-\vecc{q}^2<0.\nn
\end{gather}

From the on mass shell condition of recoil proton $(p-q)^2=M^2$, one
infers
\begin{equation} \label{eq4}
s\beta_q=-(\vecc{q}^2+M^2\alpha_q)/(1-\alpha_q)\approx -\vecc{q}^2.
\end{equation}
We use here the smallness of $M^2\alpha_q=(M^2/s)(s_1+\vecc{q}^2)$
compared with $\vecc{q}^2$. Here $s_1=(q_++q_-)^2$ -- invariant mass
square of pair, assumed to be of order $4m^2$.

For the case of large $Q$ one can put considered
$Q^2=s\alpha_q\beta_q-\vecc{q}^2$ to $Q^2=-\vecc{q}^2=-q^2$.

The ratio of transversal and longitudinal component of momentum of
recoil proton (laboratory frame implied) is
\begin{gather} \label{eq5}
\tan\theta=\frac{\vecc{p}'_\bot}{\vecc{p}'_{||}}=
\frac{|\vecc{q}|}{(\vecc{q}^2/2M)}=\frac{2M}{q}.
\end{gather}

This relation, first mentioned in paper of Benaksas and Morrison
\cite{BM}, can be written in different form in terms of the value
for 3--vector of momentum of recoil proton $P$
\begin{gather} \label{eq6}
\frac{P}{2M}=\frac{\cos\theta}{\sin^2\theta},\quad
q^2=4M^2\cot^2\theta,
\end{gather}
with $\theta$ is the angle between the directions of initial photon
and recoil proton in laboratory frame (see more exact formula in
Appendix A).

Matrix element of charged lepton or pion pair production in lowest
order of QED perturbation theory (keeping in mind BH mechanism) has
the form
\begin{gather} \label{eq7}
M^i=\frac{(4\pi\alpha)^{3/2}}{-q^2}J^p_\nu O^i_{\mu\lambda}
\vare^\lambda(k)g^{\mu\nu},\; i=l(\m{lept}),\pi(\m{Ps}),
\end{gather}
with proton current defined as
$$J^p_\nu=\bar{u}(p')[F_1(Q^2)\gamma_\nu+
\frac{[\hat{Q},\gamma_\nu]}{4M}F_2(Q^2)]u(p),$$ and $F_{1,2}$ --
proton form factors. Compton lepton tensor has the form
\begin{equation*}
O^l_{\mu\lambda}=
\bar{u}(q_-)[\gamma_\mu\frac{\hat{q}_--\hat{Q}+\!m}{D_+}\gamma_\lambda+
\gamma_\lambda \frac{\hat{Q}-\hat{q}_++\!m}{D_-}\gamma_\mu]v(q_+),
\end{equation*}
similarly Compton pion tensor
\begin{equation*}
\begin{split}
O^\pi_{\mu\lambda}&=-2g_{\mu\lambda}+
\frac{(2q_--k)_\lambda(Q-2q_+)_\mu}{D_-}\\
&+\frac{(k-2q_+)_\lambda(2q_--k)_\mu}{D_+},\quad
D_\pm=(k-q_\pm)^2-m^2.
\end{split}
\end{equation*}
These tensors obey the gauge invariance requirements
$O^i_{\mu\lambda}Q^\mu=O^i_{\mu\lambda}k^\lambda=0$.

Using Gribov prescription for Green function of virtual photon in
Feynman gauge and omitting small contributions in frames of declared
accuracy
$$g^{\mu\nu}=g_\bot^{\mu\nu}=\frac{2}{s}\big[\tilde{p}^\mu
k^\nu+\tilde{p}^\nu k^\mu\big]\approx \frac{2}{s}\tilde{p}^\mu
k^\nu,$$ one can put the matrix element, extracting explicitly the
factor $s$ in form
\begin{gather} \label{eq8}
M^i=s\frac{(4\pi\alpha)^{3/2}}{-{q}^2}N^p \frac{2}{s}\tilde{p}^\mu
e^\lambda O^i_{\mu\lambda},\;\; N^p=\frac{1}{s}J^p_\eta k^\eta,\;\;
i=l,\pi.
\end{gather}

Both light--cone projections of proton current and Compton tensors
are finite in large $s$ limit. Summing on proton spin states one has
for proton current projection square
$$ \sum |N^p|^2=2F(q^2)=2\big[F_1^2(-q^2)+
\frac{{q}^2}{4M^2}F_2^2(-q^2)\big].$$

Expressing the phase volume of the final particles in terms of
Sudakov variables \cite{BFKK}
\begin{align}  \label{eq9}
\dd\Gamma&=(2\pi)^{-5}\frac{\dd^3q_-}{2\epsilon_-}
\frac{\dd^3q_+}{2\epsilon_+}\frac{\dd^3p'}{2E'}
\delta^4(p+k-p'-q_--q_+)\\&=(2\pi)^{-5} \frac{\dd^2q \dd^2q_-\dd
x_-}{4sx_-x_+}\nn,
\end{align}
where we introduce the unit factor $\dd^4Q\delta^4(p-Q-p')$ and
besides we have used
\begin{align*}
\frac{\dd^3q_-}{2\epsilon_-}&=\dd^4q_-\delta(q_-^2-m^2)\\&=
\frac{s}{2}\dd^2q_- \dd\alpha_-\dd
x_-\delta(s\alpha_-x_--\vecc{q}_-^2-m^2).
\end{align*}

Further operations, summing over spin states of leptons of square of
matrix element, performing the integration over pair energy
fractions $x_-$, $x_+$, ($x_-+x_+=1$) and its transversal momentum
$\dd^2\vecc{q}_-$, (conservation low provides
$\vecc{q}_-+\vecc{q}_+=\vecc{q}$), and using the photon polarization
matrix $\vare_i\vare^*_j=(1/2)\big[I+ \xi_1\sigma_1+
\xi_3\sigma_3\big]_{ij}$, ($\xi_{1,3}$ -- Stokes parameters, $I$ --
unite matrix, $\sigma_i$ -- Pauli matrices) is straightforward but
tedious. The result can be written in the form
\begin{gather}  \label{eq10}
\frac{\dd\sigma_0^{\gamma\m{p}\to a^i\bar{a}^i\m{p}}}{\dd\phi
\dd\theta}=\frac{1}{2\pi}\frac{\dd\sigma^{\gamma\m{p}\to
a^i\bar{a}^i\m{p}}}{\dd\theta}\bigl(1+ \Lambda^iP_l \cos
2(\phi-\phi_1)\bigr),
\end{gather}
where $P_l=\sqrt{\xi_1^2+\xi_3^2}$. The azimuthal angle $\phi$ is
the angle between two transversal to photon direction vectors:
photon linear polarization $\vecc{\vare}$ and $\vecc{q}$; $\phi_1$
is angle between $\vecc{\vare}$ and axes $x$; $P_l$ is degree of
photon linear polarization and
\begin{align}  \label{eq11}
\frac{\dd\sigma^{\gamma\m{p}\to a^i\bar{a}^i\m{p}}_0}{\dd\theta}&=
\frac{\alpha^3}{3\pi M^2}F(q^2)\frac{\sin
\theta}{\cos^3\theta}\,a^i,\\ \nn a^l&=\frac{4L_l}{R_l}+1-
\frac{m_l^2L_l}{M^2R_l}\tan^2\theta, \\ \nn
a^\pi&=\frac{1}{2}\bigl(\frac{2L_\pi}{R_\pi}-1+
\frac{m_\pi^2L_\pi}{M^2R_\pi}\tan^2\theta\bigr),
\end{align}
$\Lambda^i$ is azimuthal asymmetry
\begin{align}  \label{eq12}
\Lambda^i=\frac{b^i}{a^i},\quad
&b^l=-(1-\frac{m_l^2L_l}{M^2R_l}\tan^2\theta),\\ \nn
&b^\pi=\frac12(1-\frac{m_\pi^2L_\pi}{M^2R_\pi}\tan^2\theta).
\end{align}
In the equations (\ref{eq11}, \ref{eq12}) quantities $L_i$, $R_i$
are
\begin{gather*}
R_i=\sqrt{1+\frac{m_i^2}{M^2}\tan^2\theta},\\
L_i=\ln(\frac{M}{m_i})+\ln\cot\theta+\ln(1+R_i).
\end{gather*}

It is interesting to consider distribution $\dd
\sigma^{\gamma\m{p}\to a^i\bar{a}^i\m{p}}/\dd q$ of recoil proton on
the value $q$. Calculations of this distribution were carried out on
the base of formula, which is obtained from (10, 11) after
substitution $\theta=\arctan(2M/q)$ (see (\ref{eq6}))
\begin{gather}
\frac{\dd\sigma^{\gamma\m{p}\to a^i\bar{a}^i\m{p}}}{\dd
q}=\frac{8\alpha}{3q^3}F({q}^2)\tilde{a}^i,\\
\nn \tilde{a}^l=\frac{4q}{\sqrt{4m_l^2+q^2}}\tilde{L}_l+1-
\frac{4m_l^2}{q\sqrt{4m_l^2+q^2}}\tilde{L}_l, \\ \nn %
\tilde{a}^\pi=
\frac12\biggl(\frac{2q}{\sqrt{4m_\pi^2+q^2}}\tilde{L}_\pi-1
+\frac{4m_\pi^2}{q\sqrt{4m_\pi^2+q^2}}\tilde{L}_\pi\biggr), \\ \nn %
\tilde{L}_i=\ln \bigl(\frac{q+\sqrt{4m_i^2+q^2}}{2m_i}\bigr),\quad
i=l,\pi.
\end{gather}

\begin{figure}[htb]
\begin{center}
\includegraphics[scale=0.6]{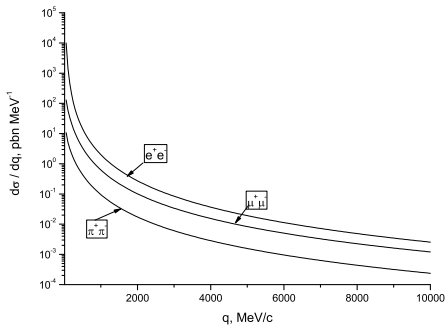}
\caption{The distributions $\dd\sigma^{i}/\dd q$ in units of pbn
MeV$^{-1}$ for the cases of $\m{e}^+\m{e}^-$ pair, $\mu^+\mu^-$ pair
and $\pi^+\pi^-$ pair production as function of $q$.} \label{fig1}
\end{center}
\end{figure}

At the Fig. \ref{fig1} the distributions $\dd \sigma^i/\dd q$ for
each of considered processes are depicted. For numerical calculation
we used the dipole approximation \cite{AR}
\begin{gather*}
F_E=\frac{F_M}{\mu}=\frac{1}{\big(1+
\frac{{q}^2[\m{GeV}^2]}{0.71^2}\big)^2}, \\
F_E=F_1-F_2\frac{q^2}{4M^2},\quad F_M=F_1+F_2.
\end{gather*}
with $\mu=2,79$ - anomalous magnetic moment of proton. Function
$F(q^2)$ in dipole approximation has the form
\begin{equation*}
F({q}^2)=\frac{4M^2+q^2\mu^2}{(4M^2+q^2)(\frac{q^2
[\m{GeV}^2]}{(0,71)^2}+1)^4}.
\end{equation*}

\begin{figure}[htb]
\begin{center}
\includegraphics[scale=.7]{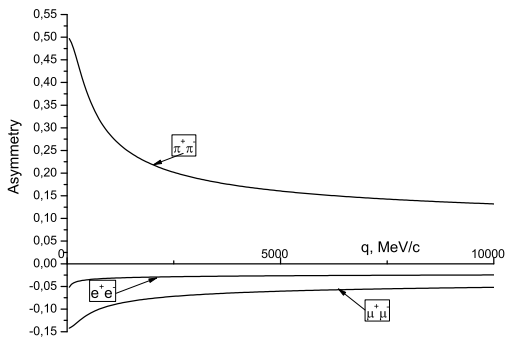}
\caption{Asymmetry $\Lambda^i$ for the cases of $\m{e}^+\m{e}^-$
pair, $\mu^+\mu^-$ pair and  $\pi^+\pi^-$ pair production as
function of $q$. } \label{fig2}
\end{center}
\end{figure}

At the Fig. \ref{fig2} the asymmetries $\Lambda^i$ as function of
momentum $q$ for each of considered processes are shown.

At the Fig. \ref{fig3} mentioned asymmetries as function of the
scattering angle $\theta$ are shown. The ratio
$$ \Lambda_{tot}=\frac{b^\m{e}+b^\mu+b^\pi}{a^\m{e}+a^\mu+a^\pi}$$
can be considered as averaged over all processes asymmetry which
estimate the total influence of initial photon linear polarization
on the value of recoil proton azimuthal asymmetry. This value is
also presented at the Fig. \ref{fig3}.

\section{Discussion}

\begin{figure}[htb]
\begin{center}
\includegraphics[scale=.75]{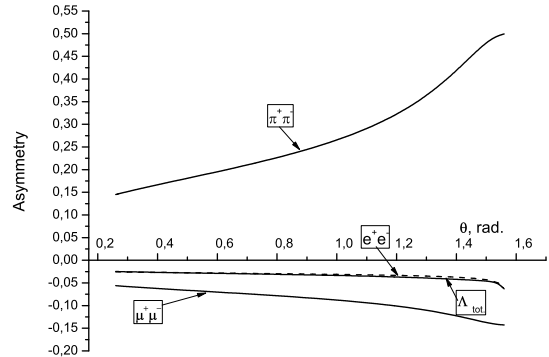}
\caption{Asymmetry $\Lambda^i$ for the cases of $\m{e}^+\m{e}^-$
pair, $\mu^+\mu^-$ pair and  $\pi^+\pi^-$ pair production and also
$\Lambda_{tot}$ as function of scattering angle $\theta$. }
\label{fig3}
\end{center}
\end{figure}

From the figures (\ref{fig2}, \ref{fig3}) one can see that in the
inclusive setup of the process of charged pairs production by
interaction of linearly polarized high energy photon with proton
distribution of recoil proton has rather essential azimuthal
asymmetry, from 0.02 at the relatively small polar angles $\theta$
up to $\Lambda_{tot}\sim0.05$ at $\theta\sim\pi/2$.

In exclusive setup for processes with more heavy particles than
e$^+$e$^-$, mentioned asymmetry increases. Particularly interesting
is the process of $\pi^+\pi^-$ pair photoproduction. One can see
that azimuthal asymmetry of recoil proton in this process reaches
the value $\Lambda^\pi\sim0.5$ at the region of small transferred
momentum or for polar angles close to the value $\theta\sim \pi/2$.
This features of $\pi^+\pi^-$ pair photoproduction process allows
one to hope that this process can be considered as the polarimetric
process. We shall discuss this process from this point of view in
the more details in the next work.

The inclusive on recoil proton distribution is the sum on all
possible channels including fermion ($\m{e}^+\m{e}^-$, $\mu^+\mu^-$,
$\tau^+,\tau^-$) and pseudoscalar meson ($\pi^+\pi^-$,
$\m{K}^+\m{K}^-$) pairs. Production of heavy resonances such as
$\rho^\pm$ meson can be excluded using experimental cuts.

The suggested method of measuring the recoil distributions can
provide the independent way to control the luminosity and
polarization properties of photon beam.

In paper \cite{VK} the photoproduction of electron--positron pair on
electron was considered in lowest order of PT. The radiative
corrections to cross section were considered in paper \cite{VKM} and
in all orders of PT on parameter $Z\alpha$ in paper \cite{IM} --
both for unpolarized case. It turns that for $Z<6$ our results can
be applied for photoproduction on nuclei with relevant modification
of $F(q^2)$. The radiative corrections can change the values $a^i$,
$b^i/a^i$ in frames of 1--2$\%$.

The proton recoil momentum measurements can as well be arranged in
$\m{ep}\to X_\m{ep'}$ and $\m{pp}\to X_\m{pp'}$ experiments with
initial proton at rest. Using the Weizs\"{a}cker--Williams
approximation the corresponding cross sections can be written as
\begin{gather}
\dd\sigma^{\m{ep}\to X_\m{e
p'}}=\frac{2\alpha}{\pi}\int\limits_{2m}^{s/(2M)}\frac{\dd\omega}
{\omega}\big[\ln\frac{s}{2\omega
m_\m{e}}-\frac{1}{2}\big]\dd\sigma^{\gamma \m{p}\to a\bar{a}\m{p'}}
\end{gather}
for electron--proton collisions and
\begin{gather}
\dd\sigma^{\m{pp}\to X_\m{p
p'}}=\frac{2\alpha}{\pi}\int\limits_{2m}^{s/(2M)}\frac{\dd\omega}
{\omega}\big[\ln\frac{s}{2M\omega}-\frac{1}{2}\big]
\dd\sigma^{\gamma \m{p}\to a\bar{a}\m{p'}},
\end{gather}
for proton--proton collisions with $s=2EM$, $E$ is the energy of
initial electron or proton and $\dd\sigma^{\gamma \m{p}\to
a\bar{a}\m{p'}}$ are given above. Inferring these formulae we
supposed that the transversal momentum of projectile (e or p) does
not exceed $M$. The polarization vector of virtual photon
interacting with proton at rest is directed along this projectile
transverse momentum.

\section{Acknowledgements}
One of us (EAK) is grateful to Institute of Physics, SAS. We are
grateful to participants of seminar of theoretical physics SAS. One
of us (Y.P) is grateful to Grant STSU-3239. The work was partly
supported also by Slovak Grant Agency for Sciences VEGA, grant No.
2/4099/26.

\appendix
\section{}
More exact formula which takes into account power corrections for
recoil proton momentum has a form \cite{BMPV}
\begin{gather}  \label{YuPcom1}
p=M\frac{(s-M^2)(s-s_1-m^2)\cos\theta\pm
(s+M^2)\sqrt{D_1}}{4M^2s+(s-M^2)^2\sin^2\theta};  \\ \nonumber
 D_1=(s-s_1+m^2)^2-4M^2s-(s-M^2)^2\sin^2\theta .
\end{gather}
Under condition $s\gg M^2$ upper branch of (\ref{YuPcom1}) passes to
\begin{gather}  \label{YuPcom2}
p=M\biggl(\frac{2\cos\theta}{\sin^2
\theta}-\frac{M^2}{s}\frac{(1+\cos^2\theta)(1+3\cos^2\theta)}{\cos\theta
\sin^4\theta}\\ \nonumber
-\frac{(s_1-4m^2)}{s}\frac{(1+\cos^2\theta)}{\cos\theta
\sin^2\theta}+O\bigl(\frac{M^4}{s^2},\frac{s_1^2}{s^2}\bigr)\biggr),
\end{gather}
with $s_1$ -- invariant mass squared of pair produced.


\begin{thebibliography}{99}
\bibitem{BM}
  D.~Benaksas and R.~J.~Morrison,
  Phys.\ Rev.\  {\bf 160} (1967) 1245.
\bibitem{BFKK}
  V.~N.~Baier, E.~A.~Kuraev, V.~S.~Fadin and V.~A.~Khoze,
  Phys.\ Rept.\  {\bf 78}, 293 (1981).
\bibitem{AR}
  A.~I.~Akhiezer, M.~P.~Rekalo, \emph{Electrodynamics of Hadrons}
  (in Russian), Naukova Dumka, Kiev, 1977.
\bibitem{VK}
  E.~A.~Vinokurov and E.~A.~Kuraev,
  Zh.\ Eksp.\ Teor.\ Fiz.\  {\bf 63}, 1142 (1972).
\bibitem{VKM}
  E.~A.~Vinokurov, E.~A.~Kuraev and N.~P.~Merenkov,
  Zh.\ Eksp.\ Teor.\ Fiz.\  {\bf 66}, 1916 (1974).
\bibitem{IM}
  D.~Ivanov and K.~Melnikov,
  Phys.\ Rev.\ D {\bf 57}, 4025 (1998)
\bibitem{BMPV}
  V.~F.~ Boldyshev, E.~A.~ Vinokurov,  N.~P.~ Merenkov and
  Yu.~P.~ Peresunko,\ Phys.\ Part.\ Nucl. {\bf 25}, 292 (1994).
\end{thebibliography}
\end{document}